\documentclass{article}
\usepackage{amsfonts}
\usepackage{graphicx}
\usepackage{epsfig}
\usepackage{eepic}
\usepackage{cite}
\usepackage{graphicx}
\usepackage{dcolumn}
\usepackage{bm}
\usepackage{times,mathptm}

\newcommand{\nc}{\newcommand}
\nc{\rnc}{\renewcommand}
\nc{\ket}[1]{\left | \, #1 \right \rangle}
\nc{\bra}[1]{\left \langle #1 \, \right |}
\nc{\proj}[1]{\ket{#1}\bra{#1}}
\rnc{\vec}{\mathbf}
\nc{\ua}{\uparrow}
\nc{\da}{\downarrow}

\nc{\braket}[2]{\langle\, #1\,|\,#2\,\rangle}
\nc{\half}{\frac{1}{2}}

\nc{\prj}{\mathcal{P}}
\nc{\hilb}{\mathcal{H}}
\nc{\pth}{\mathcal{C}}
\nc{\inprod}[2]{\braket{#1}{#2}}
\nc{\upket}{\ket{\uparrow}}
\nc{\downket}{\ket{\downarrow}}
\nc{\upbra}{\bra{\uparrow}}
\nc{\downbra}{\bra{\downarrow}}

\begin{document}
\title{On the absorption spectrum of noble gases 
at the arc spectrum limit }
\author{Ugo Fano \\
Nuovo Cimento {\bf 12}, 154-161 (1935)\\
\\
\\
A translation from the Italian original, edited by\\ Guido Pupillo, Alberto Zannoni, and Charles W. Clark}
\maketitle

This translation was undertaken to make accessible to readers of English a foundational paper on the theory of spectral line shapes, {\it Sullo spettro di assorbimento dei gas nobili presso il limite dello spettro d'arco}, U. Fano,  Nuovo Cimento, N. S. 12, 154-161 (1935).

The results of this paper are widely known via a subsequent publication by the same author in the Physical Review in 1961. \footnote{	`` Effects of Configuration Interaction on Intensities and Phase Shifts,'' U. Fano, {\it Phys. Rev.} {\bf 124}, 1866-1878 (1961)} The 1961 paper has been cited more than 4500 times, and it was judged to be among the most influential papers published in the history of the {\it Physical Review} journal series, according to a recent study that examines both numbers and time series patterns of citations. \footnote{	`` Citation Statistics From More Than a Century of Physical Review,'' S. Redner, physics/0407137 (2004)} It is, by a considerable margin, the most-cited paper that has been published under the byline of the National Bureau of Standards/National Institute of Standards and Technology (NBS/NIST). \footnote{ ``Effects of Configuration Interaction on Intensities and Phase Shifts,'' C. W. Clark, in {\it A Century of Excellence in Measurements, Standards, and Technology; A Chronicle of Selected Publications of NBS/NIST, 1901-2000}, ed. David R. Lide, National Institute of Standards and Technology SP 958 (U.S. Government Printing Office, Washington, DC, 2001), pp. 116-119.  Online: http://nvl.nist.gov/pub/nistpubs/sp958-lide/116-119.pdf }

The 1935 paper translated here lacks the generality of the 1961 paper, but its results are identical for an important limiting case, and it should be considered to be the first paper which correctly elucidates the general form of line shapes encountered in the excitation of many important atomic and condensed-matter systems.  In particular, it treats the case in which a discrete state coexists in the same energy region as a continuum of states, and accounts for the interaction between the discrete and continuum states, and the interference between their separate excitation amplitudes.  The key line-shape formula derived in the 1935 paper is identical in a practical sense to that of the 1961 paper, which is now famous as the {\it Fano profile}: it does not include a shift in the discrete-state energy due to its interaction with the continuum (as does the 1961 paper), but this is not a direct observable.

In addition to its historic interest, the 1935 paper presents its subject in a remarkably clear way, no doubt reflecting the influence of Enrico Fermi, who was Ugo Fano's supervisor at the time.  It does not use the somewhat formidable mathematical apparatus of the 1961 paper, and it offers insights which may still seem fresh even to those familiar with the subject matter (for example, Fano's observation of how a discrete state with zero excitation amplitude can cause the total excitation probability to vanish at its own energy).\\

Note on the text: the original publication does not identify equations by number.  Equation numbers have been added in the translation for readers' convenience.\\

The editors are grateful to the Societ\`a Italiana di Fisica for permission to publish this translation, and to Ms. Susan Makar, of the library of the National Institute of Standards and Technology, for much helpful assistance. \\
\\
\\
\\
\\
\\
\\
\\
\\
\\
\\
\\
\\
\\
\\
\\
\\
\\
\\
\\
\\
\\
\\
\\
\\
\\
\\
\\
\\
\\
\\
\\
\\
\\



\begin{abstract}
 Rydberg spectral lines of an atom are sometimes superimposed on the continuous spectrum of a different configuration. Effects of interaction among different configurations in one of these cases are theoretically investigated, and a formula is obtained that describes the behavior of absorption spectrum intensity. This offers qualitative justification of some experimental results obtained by BEUTLER in studies of absorption arc spectra of noble gases and $I^b$ spectra of some metal vapors.
\end{abstract}

    It is experimentally known that arc spectrum series of noble gases do not converge toward a single limit, but toward two distinct limits. The explanation is that removal of the optical electron from a noble gas atom yields an ion 
whose ground configuration does not consist of a single level, but rather a doublet $^2 P_{3/2}^o$,$^2 P_{1/2}^o$. The interval between the doublet's levels is about 1500 wave numbers for A, 5000 for Kr, and 10000 for Xe. Within this interval, two different kinds of arc spectrum terms can occur: a) continuous spectrum terms represented by the formula $ (p^5)_{3/2} $+{\it free electron}; b) discrete spectrum terms represented by the ({\it jj} coupling) formula $ (p^5)_{1/2}nl $; the latter belong to series that converge toward the $ ^2 P_{1/2}^o$ limit. In a recent work (\cite{BEUTLER1}) BEUTLER investigated absorption spectra of noble gases, obtaining the following results. At very low pressure of the noble gas (0.002 mm), continuous absorption with a regular behavior is observed for frequencies greater than the $^2 P_{1/2}^o$ limit, and also continuous absorption with characteristic intensity modulations is observed between the $^2 P_{3/2}^o$ and $^2 P_{1/2}^o$ limits.  Absorption between the two limits shows maxima that can be classified into two groups: a more peaked, and a much less peaked one; positions belonging to each of these two groups are Rydberg series that converge to the $^2 P_{1/2}^o$ limit. With increasing noble gas pressure, absorption peaks grow in intensity and width until they overlap. At the pressure of 0.030 mm the absorption is already continuous and homogeneous, starting from the $^2 P_{3/2}^o$ limit. In any case absorption due to energy levels below the $^2 P_{3/2}^o$ limit is smaller in magnitude than absorption above the same limit. The intensity distribution in the absorption spectrum is shown by BEUTLER in a graph whose characteristic appearance is reproduced in figure 1.
 He interprets single maxima as lines of the discrete spectrum, much broadened due to the large probability of self-ionization (AUGER effect) $(p^5)_{1/2}nl \rightarrow (p^5)_{3/2}$ + {\it free electron}. Wide maxima are assigned to the series $(p^5)_{1/2}nd $, and narrow maxima to the series $(p^5)_{1/2}ns $. 
    The aim of the present work is to show how it is possible to justify such an intensity distribution in a qualitative way, by supposing that positions of discrete terms do not correspond to absorption maxima, but to points located along the steep parts of the curve, which are therefore slightly shifted with respect to the former.  

     The intensity distribution in the part of the spectrum of interest is obtained by evaluating the squares of dipole matrix elements referring to transitions from the ground state to states whose energy lies between the $^2 P_{3/2}^o$ and $^2 P_{1/2}^o$ levels of the ion. If we performed this calculation starting with zeroth-order eigenfunctions, corresponding to single electronic configurations, we would find that absorption is due to superposition of a continuum of almost constant intensity with lines belonging to series that converge to the $^2 P_{1/2}^o$ limit.
\begin{figure}
\begin{center}
\includegraphics[width=1.0\linewidth]{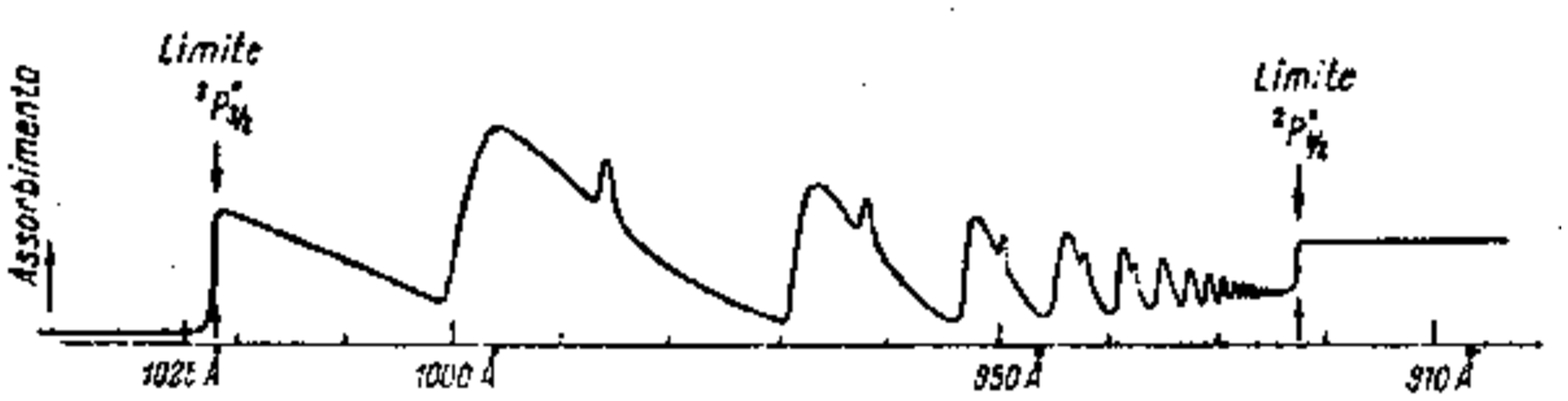}
\caption{}
(From ``Zeitschrift f\"ur Physik'', {\bf 93}, 181, 1935)
\end{center}
\end{figure}

    In order to obtain a result in agreement with BEUTLER's experimental data we should start instead with better approximate eigenfunctions, which take into account interaction between different configurations. Ordinarily, eigenfunctions of this type are obtained via perturbation theory; this method is not applicable to our case, as we deal with states belonging to the continuum whose energies are infinitely close to, and also coincident with, the energies of the discrete states. We therefore have to abandon perturbation theory and look directly for eigenfunctions of the SCHR\"ODINGER equation. We may assume as a first approximation that states which are not close to each other do not appreciably interact. Therefore, absorption in proximity of the position of a particular discrete term is obtained considering the interaction only between the term itself and the continuum.

   With the problem so outlined, it will be convenient to treat the atom as embedded in a sphere of very large radius $R$, in order to simplify the treatment of the continuum. The continuous spectrum is then replaced by a discrete one. The interval $\tau$ between two consecutive eigenvalues is almost constant for small energy variations, and is inversely proportional to $R$. The corresponding eigenfunctions contain a factor related to the free electron, which far from the atom takes the form

\begin{equation}
\frac{1}{r} \sin \left( \sqrt{\frac{8 \pi^2 m}{h^2} W} r + \delta \right) f (\theta \phi) \nonumber
\end{equation}

\noindent so that the normalization coefficient in the $R \rightarrow \infty $ limit is proportional to $1 / \sqrt{R}$. It follows that if we let $R$ go to $\infty$ we have to express a factor of $\sqrt{\tau}$ in the normalization coefficient. By invoking the fact that states with very different energies interact weakly, we can finally consider the spectrum of the $(p^5)_{3/2}+ ${\it free electron} configuration to be produced by the succession of eigenvalues

\begin{equation}
E_n=n \tau ,\ \ \ \ \ \ \ \ \ \ \ \ \ \ \  (-\infty < n < \infty) \nonumber
\end{equation}

\noindent where the energy of the discrete state under consideration is defined as the zero of energy. 

Let $\phi$ be the zeroth-order eigenfunction corresponding to the discrete term, and $\psi_n$ the one corresponding to the eigenvalue $E_n$. From the above hypothesis it follows that a perturbed eigenfunction whose energy is close to the discrete term must have the form:

\begin{equation}
\psi=\sum^{\infty}_{n=-\infty}a_n \psi_n + b \phi \nonumber
\end {equation}

Let $V$ be the interaction between electrons (which is mainly electrostatic), and define the first-order approximate energy as the energy associated with a given configuration (the sum of the eigenvalue of the equation for independent electrons, the exchange energy, and the diagonal term of $V$). The SCHR\"ODINGER equation for $\psi$ is thus decomposed into the infinite system of equations:

\begin{equation}
E a_n = E_n a_n + b V_n ,\ \ \ \ \  E b = \sum a_n V_n  \nonumber
\end{equation}

\noindent where $V_n = ( \phi | V | \psi_n)$ is supposed to be real, for the sake of simplicity.

Let us now introduce a new hypothesis, that is, $\psi_n$ is independent of $n$ at distances from the origin of the order of the atomic radius. It follows that $V_n=q$ is constant, and the last equation reads:

\begin{equation}
\frac{E b}{q}=\sum a_n  \nonumber
\end{equation}

\noindent while the other equations give:

\begin{equation}
a_n=\frac{b q}{E-E_n} \nonumber
\end{equation}

Substituting, we obtain:

\begin{equation}
(*) \ \ \ \ \ \ \   E=q^2 \sum \frac{1}{E-E_n}=q^2 \sum \frac{1}{E-n \tau}=\frac{q^2 \pi}{\tau} \cot \frac{E \pi}{\tau} \nonumber
\end{equation}

\noindent which determines the eigenvalues. In order to find $b$, we impose the following normalization condition (where $dv$ is the element of volume in configuration space):

\begin{eqnarray}
1 &=& \int |\psi|^2 dv =\sum a_n^2+b^2 = b^2 \left\{ 1 + \sum \frac{q^2}{(E-E_n)^2} \right\} \nonumber\\
&=& b^2 \left\{ 1 - \frac{\partial}{\partial E} \sum \frac{q^2}{E-E_n} \right\}
= b^2 \left\{1+\frac{q^2 \pi^2}{\tau^2} \frac{1}{\sin^2\frac{E \pi}{\tau}} \right\} \nonumber\\
&=& b^2 \left\{1+\frac{q^2 \pi^2}{\tau^2} \left( 1 + \frac{E^2 \tau^2}{q^4 \pi^2}\right)  \right\}  \nonumber\\
\end{eqnarray}

\noindent therefore

\begin{equation}
\psi=\frac{\phi + \sum \frac{q}{E-E_n} \psi_n}{\sqrt{1+\frac{q^2 \pi^2}{\tau^2}+\frac{E^2}{q^2}}} \nonumber\\
\end{equation}

Let us consider $X_n=( u | x | \psi_n )=X_c$ to be independent of $n$, where $u$ is the ground state eigenfunction, and $X_0=(u |x| \phi)$; the square of the $x$-component of the dipole matrix element is, taking (*) into account:

\begin{equation}
X^2=\frac{\left\{ X_0 + \frac{E}{q} X_c \right\}^2 }{1+\frac{q^2 \pi^2}{\tau^2}+\frac{E^2}{q^2}}=\frac{\left\{ X_c + \frac{q}{E} X_0 \right\}^2 }{1+\frac{q^2 }{E^2}+\frac{q^4 \pi^2}{E^2 \tau^2}} \nonumber
\end{equation} 

Having obtained this result, we have to take the limit $R \rightarrow \infty $. Matrix elements $X$, $X_c$ and $q$ contain a factor of $\sqrt{\tau}$, because of their definition; it is therefore convenient to set $X=\bar{X}\sqrt{\tau}$, $X_c=\bar{X}_c\sqrt{\tau}$, $q=\bar{q}\sqrt{\tau}$. Actually the quantity we are interested in is $\bar{X}^2$, since $( \bar{X}^2+\bar{Y}^2+\bar{Z}^2 )dE$ determines the transition probability from the ground state to a state of energy in the range $dE$. Therefore, we have:

\begin{equation}
\bar{X}^2 \tau=\frac{\left\{ \bar{X}_c +\frac{\bar{q}}{E} X_0 \right\}^2 }{1+\frac{\bar{q}^2 \tau}{E^2}+\frac{\bar{q}^4 \pi^2}{E^2}}\tau  \nonumber
\end{equation}

\noindent and taking the limit $\tau \rightarrow 0$

\begin{equation}
\bar{X}^2 =\frac{\left\{ \bar{X}_c +\frac{\bar{q}}{E} X_0 \right\}^2 }{1+\frac{\bar{q}^4 \pi^2}{E^2}} \nonumber\\
\end{equation}

The same formulae hold for $\bar{Y}^2$ and $\bar{Z}^2$. Setting $D=(\bar{X},\bar{Y},\bar{Z})$, $D_c=(\bar{X}_c,\bar{Y}_c,\bar{Z}_c)$, $D_0=(X_0,Y_0,Z_0)$, we obtain:

\begin{equation}
|D|^2 = \frac{| D_c + \frac{\bar{q}}{E} D_0  |^2}{1+\frac{\bar{q}^4 \pi^2}{E^2}} =\frac{| D_c |^2}{1+\frac{\bar{q}^4 \pi^2}{E^2}} + \frac{\bar{q}^2 | D_0 |^2}{E^2+\bar{q}^4 \pi^2} + \frac{2 E \bar{q}}{E^2+\bar{q}^4 \pi^2} D_c \times D_0 \nonumber
\end{equation}

In a small enough range of frequencies, this quantity can be regarded as proportional to the absorption intensity.

In order to discuss this formula, it is convenient to examine the three-term expansion. The first two terms have a well defined physical meaning, since $| D |^2$ reduces to them if $D_0$ or $D_c$ goes to zero, respectively. Therefore we observe that if the dipole matrix element associated with the continuous spectrum is zero, we obtain as the absorption spectrum a line which is broadened due to the AUGER effect, and whose width is given by $\nu = \frac{2 \pi \bar{q}^2}{h}$, as expected. If instead the matrix element associated with the discrete state is zero, the state itself affects absorption by the continuum, in that the latter vanishes at the position of the former. The third term is truly distinctive, as it results in net absorption being not simply obtained as a superposition of absorptions due to discrete and continuum terms, albeit mutually influencing each other; this term represents a shift of absorption intensity, or, in other words, it diminishes the intensity on one side of the discrete term's position, and increases it by the same amount on the other side. 
\begin{figure}
\begin{center}
\includegraphics[width=1.0\linewidth]{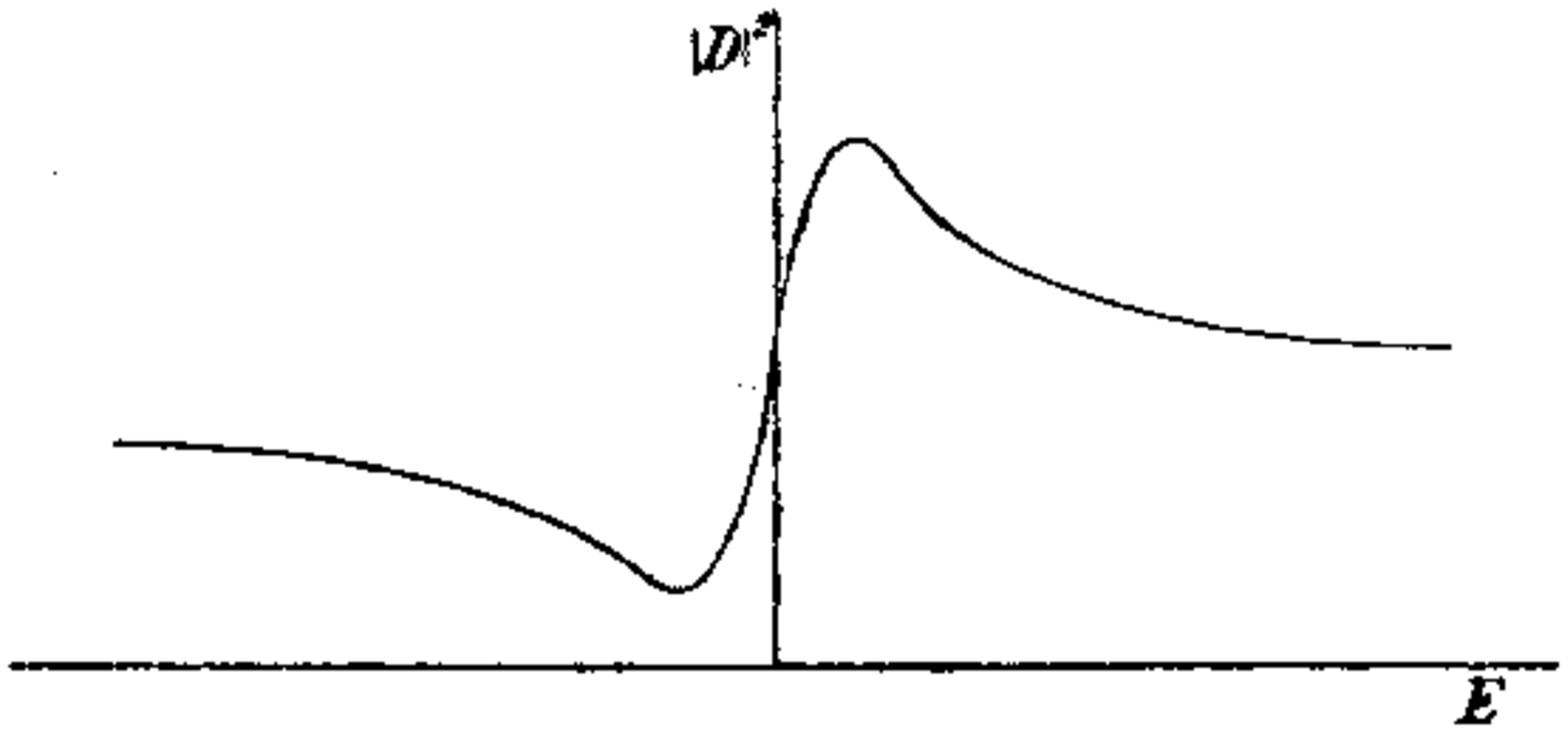}
\caption{}
$|D_c|=2$ ; $|D_0|=4.2$ ; $\bar{q}=0.6$ ; $D_c \times D_0 =6$ ; $\frac{|D_0|^2}{\bar{q}^2 \pi^2}=5$ arbitrary units
\end{center}
\end{figure}

In figure 2, $| D |^2$ is depicted as a function of $E$ for some values of $| D_0 |$, $|D_c|$, $D_0 \times D_c$, $\bar{q}$ (the parameters that determine the curve); values have been chosen to show that {\it the curve} itself {\it can have a behavior that justifies theoretically the results obtained by Beutler}. Characteristic features of the curve that readily result from the discussion of its equation are: $a$) the curve goes asymptotically to $|D_c|^2$ for $E  \rightarrow \pm \infty $; $b$) the ordinate of the intersection of the curve with the $E=0$ axis depends only on $| D_0 |$ and $\bar{q}$, as it is equal to $\frac{| D_0 |^2}{\bar{q}^2 \pi^2}$; $c$) the curve has a maximum and a minimum on opposite sides of the $E=0$ axis; in particular, if $D_c$ is parallel to $D_0 $, the mimimum is equal to zero; $d$) the difference between the abscissa of the maximum and of the minimum is of the order of $\bar{q}^2 \approx \frac{h}{2 \pi \tau} $, where $\tau$ is the lifetime of the discrete term with respect to the AUGER effect (estimating $\tau$ as $10^{-14}$ sec, one gets $\bar{q}^2/hc \sim 500$ wave numbers).

Obviously, due to the simplifying assumptions that we used, the result obtained has merely a qualitative value, which is to show the behavior of the curve. 

That the derived formula fails to fulfill the sum rule is to be attributed to the hypothesis adopted, since it indeed should yield:

\begin{equation}
\lim_{U \rightarrow \infty} \left\{ \int_{-U}^U | D |^2 dE - 2 U | D_c |^2 - |D_0 |^2 \right\} = 0. \nonumber\\
\end{equation}

In fact, we assumed the presence of a continuous spectrum of infinite extent, with $ | D_c | $ constant, which is physical nonsense as it would result in an infinite number of dispersion electrons. This incorrectness is particularly evident in the limiting case $D_0 = 0$, where it appears that the number of dispersion electrons of the continuum is reduced by a factor of $\frac{1}{1+\frac{\bar{q}^4 \pi^2}{E}}$ in the vicinity of the perturbing discrete term, without being correspondingly increased in other parts of the spectrum, so that the total sum of dispersion electrons does not change.

A trial calculation has shown that the derived formula is not even susceptible to a rough numerical evaluation, due to the large number of electrons that must be included for noble gases, and the poor approximation achievable in evaluating individual integrals.
\begin{figure}[t]
\begin{center}
\includegraphics[width=1.0\linewidth]{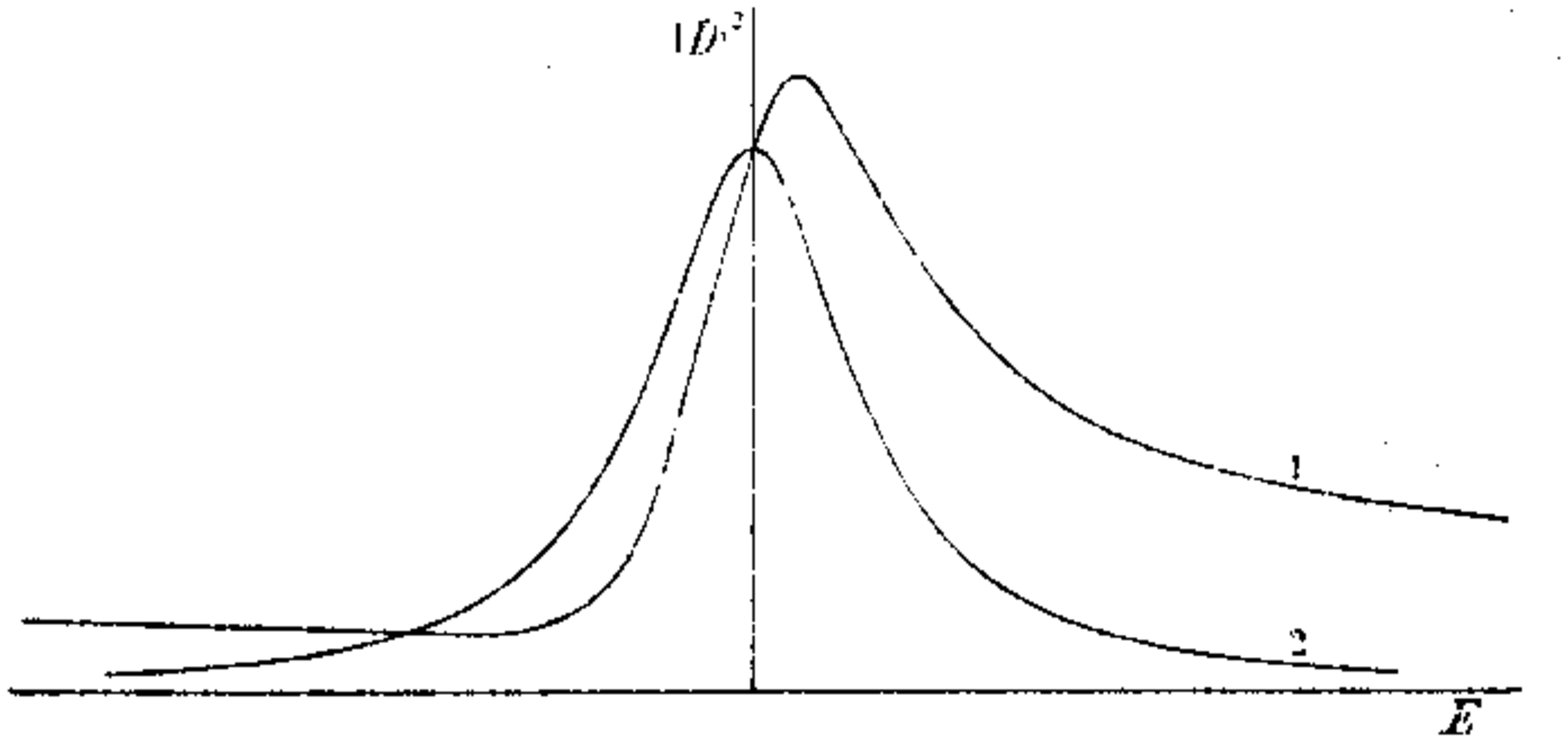}
\caption{}
Curve 1: $|D_c|=1$ ; $|D_0|=4.2$ ; $\bar{q}=0.6$ ; $D_c \times D_0 =3$ ; $\frac{|D_0|^2}{\bar{q}^2 \pi^2}=5$ \\
Curve 2: $|D_c|=0$ ; $|D_0|=4.2$ ; $\bar{q}=0.6$ ; $D_c \times D_0 =0$ ; $\frac{|D_0|^2}{\bar{q}^2 \pi^2 } = 5$ \\
arbitrary units 
\end{center}
\end{figure}

{\it Application of the obtained formula to line broadening phenomena in $I^b$ absorption spectra.} - $I^b$ spectra, obtained by excitation of an electron belonging to the outermost closed shell, have been studied by BEUTLER in a series of important works (\cite{BEUTLER2}). Superposition of discrete terms of $I^b$ spectra upon continuum terms of ordinary arc spectra gives rise to the same situation that occurs in noble gases' spectra between the two limits of the arc spectrum. Nevertheless, the phenomena look different, since, up to the present, in known cases absorption due to the continuum: $a$) is much less intense than absorption due to discrete terms of $I^b$ spectra, and $b$) decreases very rapidly in intensity with increasing frequency.

BEUTLER observed that some $I^b$ series have bright and narrow lines superimposed upon the continuous spectrum. He explains this phenomenon by showing that interaction among terms belonging to these series and to the continuum vanishes. On the other hand, other series have diffuse lines, which he describes as asymmetrically broadened. In every diffuse series, lines tend to become symmetric again and to narrow as frequency increases, as the intensity of the continuum on which they are superimposed decreases.

In figure 3 two different graphs of $ | D |^2 $ as a function of $E$ are shown; curve 1 is obtained taking a value of $ | D_c |^2 $ small compared to $ \frac{| D_0 |^2}{\bar{q}^2 \pi^2} $, curve 2 is obtained for the same values of $ | D_0 |$ and $\bar{q}$, and with $D_c = 0$. Line shape peculiarities in diffuse series of $I^b$ spectra can therefore be explained, since the data relative to lower frequency lines are those utilized to obtain figure 1, while by increasing frequency we get closer to conditions corresponding to figure 2. Narrowing of the lines with increasing frequency is probably due to the fact that the $\bar{q}$ interaction tends in general to decrease with increasing total quantum number of the corresponding discrete term.\\
\\
\\
I want to deeply thank Prof. FERMI who guided and helped me throughout this work. 
{\it Rome, Istituto di Fisica della R. Universit\`a, February 1935-XIII.}
 


\begin{thebibliography}{99}


\bibitem{BEUTLER1} BEUTLER, Zs. f\"ur Physik, {\bf 93}, 177, 1935

\bibitem{BEUTLER2} Zs. f\"ur Physik, {\bf 86},495,710; {\bf 87},19,176,187; {\bf 88},25;{\bf 91},143,202,218.

\end{thebibliography}
\end{document}